\documentclass[a4paper, conference, 10pt]{IEEEtran}
\ifCLASSINFOpdf
\else
\fi

\usepackage{tabularx,ragged2e,booktabs,caption,cite,amsmath,amssymb,latexsym,float}
\usepackage[final]{graphicx}
\usepackage{epsfig}
\usepackage{subfigure,epstopdf,array,algorithm,algpseudocode,subfigure}
\usepackage{url}
\begin{document}
%
\title{Survivability Improvement Against Earthquakes in Backbone Optical Networks Using Actual Seismic Zone Information}



%
\author{\IEEEauthorblockN{Anuj Agrawal\IEEEauthorrefmark{1},
Purva Sharma\IEEEauthorrefmark{2},
Vimal Bhatia\IEEEauthorrefmark{1},
and
Shashi Prakash\IEEEauthorrefmark{2}}
\IEEEauthorblockA{\IEEEauthorrefmark{1}Discipline of Electrical Engineering\\
Indian Institute of Technology Indore, India
\\ Email: phd1501202003@iiti.ac.in, vbhatia@iiti.ac.in}
\IEEEauthorblockA{\IEEEauthorrefmark{2}Institute of Engineering and Technology\\
Devi Ahilya University, Indore, India\\
Email: purvasharma478@gmail.com, sprakash\_davv@rediffmail.com}
}


\maketitle

\begin{abstract}
Optical backbone networks carry a huge amount of bandwidth and serve as a key enabling technology to provide telecommunication connectivity across the world. Hence, in events of network component (node/link) failures, communication networks may suffer from huge amount of bandwidth loss and service disruptions. Natural disasters such as earthquakes, hurricanes, tornadoes, etc., occur at different places around the world, causing severe communication service disruptions due to network component failures. Most of the previous works on optical network survivability assume that the failures are going to occur in future, and the network is made survivable to ensure connectivity in events of failures. With the advancements in seismology, the predictions of earthquakes are becoming more accurate. Earthquakes have been a major cause of telecommunication service disruption in the past. Hence, the information provided by the meteorological departments and other similar agencies of different countries may be helpful in designing networks that are more robust against earthquakes. In this work, we consider the actual information provided by the Indian meteorological department (IMD) on seismic zones, and  earthquakes occurred in the past in India, and propose a scheme to improve the survivability of the existing Indian optical network through minute changes in network topology. Simulations show significant improvement in the network survivability can be achieved using the proposed scheme in events of earthquakes.

\end{abstract}


%
\IEEEpeerreviewmaketitle

\section{Introduction}
Optical networks are a key technology to fulfil the exponentially increasing bandwidth demand due to various emerging applications such as cloud computing, bandwidth on demand, etc. Wavelength division multiplexing (WDM) \cite{zang2000review} has emerged as the most popular and widely used technology for optical networks. WDM divides the optical C-band spectrum into fixed wavelength channels, usually of $50$ GHz, where each channel/wavelength can carry a lightpath and multiple wavelengths can be transmitted simultaneously through a single optical fiber. Newly introduced elastic optical networks (EONs) have been presented in the literature \cite{jinno2009spectrum} as a promising candidate to replace the widely used WDM technology. EON is advantageous to WDM network in the sense that it can adapt to the actual traffic needs and transmission reaches in contrast to the fixed wavelength grids in a WDM network. This adaptability of EON increases the traffic carrying capacity of EON much more than that of WDM networks. With the increasing traffic carrying capacity of EON, the bandwidth loss that occur during network component (node/link) failures is also greater in EONs. Hence, more robust networks need to be designed with a view to maintain high connectivity in the current and future optical networks even during multiple failures that may occur during natural disasters. The single and double link or node failures may occur in terrestrial backbone networks due to human activities, and other normal excavations. However, the effects of natural calamities such as earthquakes are highly disastrous, that may result in multiple link or node failures, consequently disturbing the network connectivity to a greater extent, and may isolate the affected country from the rest of the world. Examples include Wenchuan earthquake in China which damaged $28,765$ km sheath of optic cables\cite{wenchuan}, $2011$ Japan Tsunami\cite{japanMagazine2016}, Hurricane Sandy, among others. Similar disasters occur every few years around the globe\cite{wenchuan}. Earthquakes happen all the time around the world and the occurrence of natural disasters is increasing due to global warming and other such effects\cite{asimakopoulou2011collective}. Hence, sufficient attempts should be made to utilize the seismic information made available by various meteorological agencies of the world to make prevention policies in order to minimize the economic loss, and service disruptions in events of natural disasters. In this work, we propose a node relocation method in backbone optical networks using the actual seismic zone information of India\cite{iitk4}. In the proposed method, the nodes of backbone optical networks have been relocated in very small area of about $260\times260$ km so that the connectivity reach of the already existing network to different parts of the country is not affected.

\begin{figure}[!t]
\centering
\includegraphics[width=3.2in]{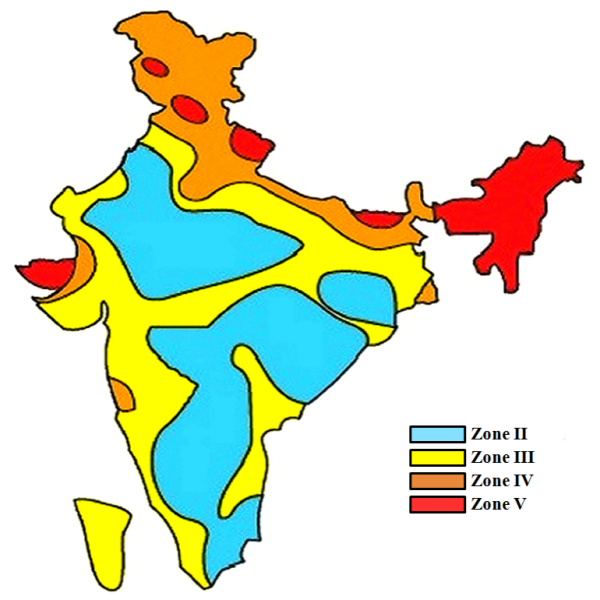}
\caption{Indian seismic zone map of 2002.}
\label{fig_sim1}
\end{figure}

The map of India is subdivided into four seismic zones zones, namely, zone II, zone III, zone IV, and zone V with the expected maximum modified mercalli (MM) intensity of seismic shaking in these zones as VI or less, VII, VIII, and IX or above, respectively, as shown in Fig. 1\cite{iitk4}\cite{nptel}. The seismic hazard maps of India have been revised from time to time since $1935$ when the geological survey of India (GSI) came up with the first national seismic hazard map of India. The seismic zone map shown in Fig. 1 is the recent version revised in $2002$ by the combined efforts of Indian meteorological department (IMD), bureau of Indian standards (BIS), GSI, and other supporting organizations. Natural calamities are hard to forecast, however, the accuracy of this $2002$ seismic zone map of India can be verified from Fig. 2 that shows the earthquakes occurred in past two years in India. Fig. 2 has been obtained using the earthquake data made available by the IMD \cite{imddaily}. All the earthquakes that occurred in India with a magnitude of $4.5$ (categorized as ``rather strong" by IMD) or more on Richter scale in India from November $17$, $2014$ to October $17$, $2016$ have been placed in Fig. 2, and an analysis based on these earthquakes has been made later in this paper. We also take into account the past major earthquakes\cite{iitk4}\cite{niceemajor} occurred in India that had a devastating effect on the nation, as shown in Fig. 3. We take RailTel optical WDM network topology \cite{pandya2014RailTelJLT}, shown in Fig.4, as a reference to study the effects of major previous earthquakes, and the earthquakes that occurred in past two years in India. With an objective to improve the network survivability using the proposed SZANR, we relocate the nodes of RailTel network topology using the seismic zone information of Fig. 1, such that the new positions of nodes lie in a nearby low seismic zone. The node relocation is performed with three constraints, namely, link length constraint (LLC), node search space constraint (NSSC), and limited movement of border nodes constraint (LMBC), which ensure that the functioning of the already existing network is not affected. The proposed scheme and constraints are discussed in detail in Section III of this paper. Section II focus on the work done in the related area. Simulation results are given in Section IV, and finally conclusions are drawn in Section V.

%

\section{Related Work}

A number of attempts have been made to improve the optical network survivability against single and multiple link or node failures through protection and restoration\cite{sahas2003survivable}. Protection is a proactive scheme in which resources are reserved in advance to maintain connectivity in events of failures. In restoration, resources are searched dynamically after failures. However, few research attempts have been made to make backbone networks robust against natural disasters, that had been quite destructive in the past for the communication networks due to multiple component failures as discussed in Section I of this paper. In\cite{habib2013disaster}, a survey of disaster survivability in optical communication networks has been presented, in which a classification of disaster based on their characteristics is given and their impact on communication networks is studied. The concept of disaster-free network has been given in \cite{saito2015concept}, which suggests that a network should be made as robust as possible against natural disasters to avoid huge economic and connectivity loss since we cannot always forecast and avoid disasters. A comprehensive study on the effects of Wenchuan earthquake of China on communication networks has been made in \cite{wenchuan}, which strongly suggests enhancement of emergency communication capabilities of network under such disasters. In \cite{dikbiyik2014proactive}, proactive and reactive methods to deal with disaster failures in optical backbone networks have been proposed with an objective to reduce penalties in case of disasters, where the study assumes probabilistic risk models to analyze the effects of earthquakes. However, in the proposed method, we consider the actual seismic information, and the improvement in survivability is achieved using minute changes in the existing real world network topology. A discussion on the nature of possible disruptions in communication networks in events of disasters, and to prepare the network and cloud services against these disasters has been presented in\cite{mukherjee2014adaptability}. A scheme to deploy new disaster aware submarine fiber cables robust against earthquakes, has been presented in\cite{msongaleli2016submarine}, based on a probabilistic model. Geographical route design for deploying new physical networks using the actual seismic hazard information of Japan has been proposed in \cite{japanMagazine2016}. Improvement in network robustness against earthquake disasters has been achieved in \cite{nov2016JLT} using additional physical fiber link deployment, however, using the proposed SZANR scheme, we achieve significant improvement in network survivability against earthquake disasters without additional deployment of any network component.

From the above discussion, it can be seen that most of the studies in the related area focus on the new network topology designing methods in order to deal with future disasters. Some of the works discuss the post-disaster measures to restore the network connectivity. Survivability can also be achieved using the conventional methods of protection and restoration, however, these methods are not efficient for severe network disruptions caused by earthquakes. Designing and deploying new physical networks or parts of it may improve the network survivability against disasters to a greater extent, however, it involves a huge capital expenditure too. In this work, we propose a scheme, in which through minute changes in the locations of nodes of already deployed network, significant improvements in network survivability can be achieved. Although the probability of failures of nodes located in a high seismic zone is low \cite{japanMagazine2016}, all the links connected to it may fail in event of an earthquake with epicenter near the node location, consequently resulting in complete connectivity loss in the affected area. Thus, in this work, we demonstrate through simulations, the effect of node relocation on the survivability of the presently deployed WDM network and future EONs against earthquakes.

\section{Seismic zone aware node relocation (SZANR)}

\begin{figure}[!t]
\centering
\includegraphics[width=3.2in]{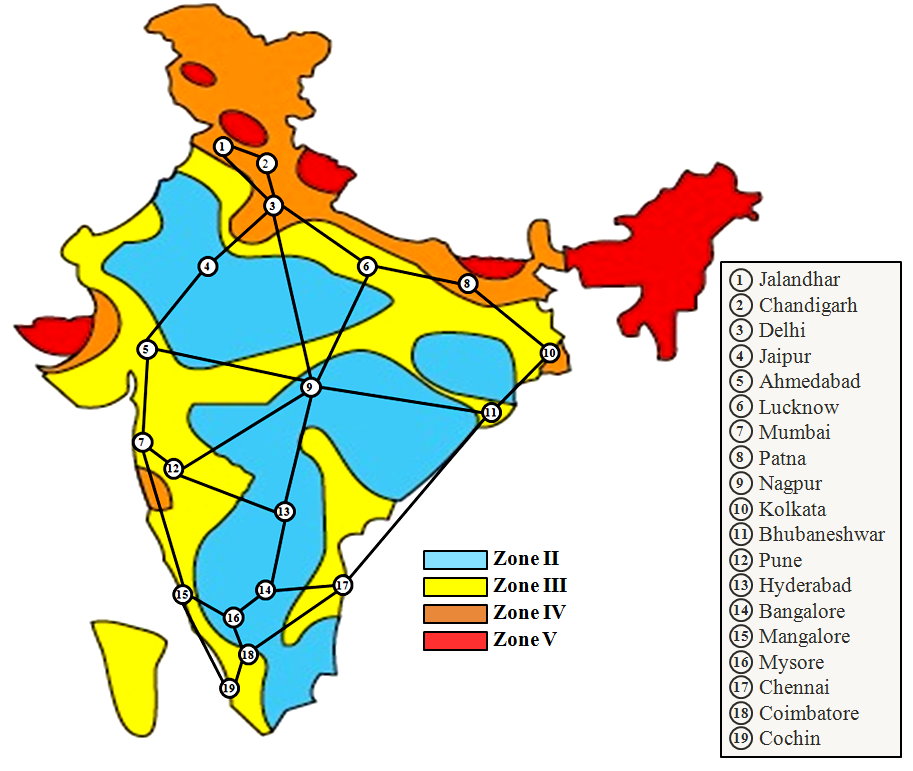}
\caption{RailTel optical network topology.}
\label{fig_sim}
\end{figure}

In the proposed SZANR scheme, we consider the actual seismic zone map of India as shown in Fig. 1. The existing RailTel network topology, as shown in Fig. 4, has been analyzed and the nodes lying in the high seismic zones are searched for relocation in a nearby low seismic zone. For node relocations, a search is performed in the two dimensional search space surrounding the original location of node. We consider an area of about $3240\times3240$ km as shown in Fig. 5. To perform node relocation search, the whole area is subdivided into a $60\times60$ grid, where each square in the grid represents an area of about $52\times52$ km. The nodes of original RailTel topology are then placed in the grid, and the grid is modelled as a matrix with the matrix elements representing the seismic zone information and the node locations. Relocation search is then performed using a simulation program developed in MATLAB environment. The relocation for all the nodes is subject to the following three constraints, viz., LLC, NSSC, and LMBC. According to LLC, the total length of the links present in the relocated node network topology must be less than or equal to the total link length of the original network topology. LLC ensures that the new relocated network topology does not require additional lengths of the fiber links. However, after node relocation, connectivity to the new relocated nodes may be achieved through additional expenditure on fiber link deployment from the original node location to the new node location. This, however, require link deployment of comparatively lesser lengths ranging from tens to hundreds of km than the deployment of new disaster aware network topologies that may require deployment of links up to lakhs of km. As per NSSC, nodes need to be relocated in the nearby area of the location of the original node only. NSSC ensures that the connectivity reach of the existing network topology is not affected and high speed communication through backbone optical network is made available to different parts of the country. Through LMBC, we ensure that the nodes located on the periphery of the map should not be moved inside, however, these nodes may be relocated if another location in nearby low seismic zone can be found on the periphery of the map, since these nodes are used to establish connections with other parts of the globe through submarine fiber cables\cite{msongaleli2016submarine}.

\begin{figure}[!t]
\centering
\includegraphics[width=3.2in]{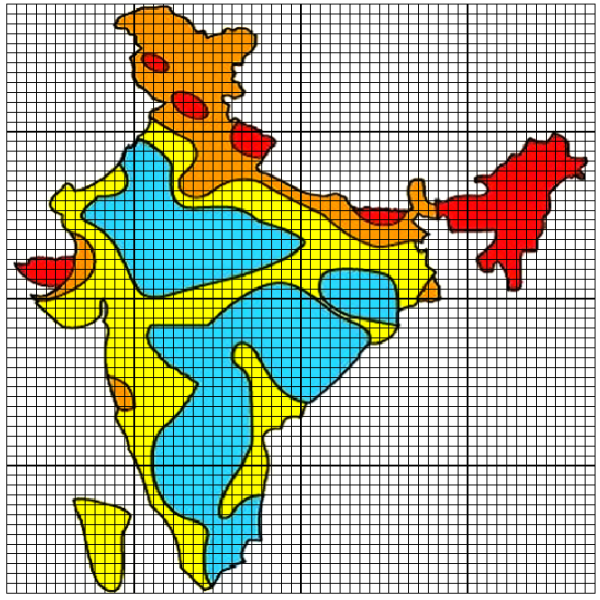}
\caption{Seismic zone map subdivided into grids.}
\label{fig_sim}
\end{figure}

\section{Simulation results and discussion}

To assess the performance of the existing optical network with that of the network topology obtained after node relocation using the proposed SZANR scheme under earthquakes, we simulate the actual earthquakes occurred in India in the past, on both network topologies. We consider the past major earthquakes that occurred since $1819$, and the earthquakes that occurred in past two years in India. The effects of the earthquakes in a particular location depend on various factors such as local variation in soil type \cite{iitk4}, water bodies, magnitude and epicenter of an earthquake, among others. To apply the effects of past earthquakes on the original and relocated network topology, we consider the historical records of past earthquakes \cite{niceemajor}, and the information publicly made available by IMD\cite{imddaily}. All the earthquakes with magnitude greater than $4.5$ on Richter scale (classified as ``rather strong" by IMD) have been considered for simulations. The effect of an earthquake decreases exponentially with distance from its epicenter\cite{japanMagazine2016}. However, the actual effect of earthquakes depend not only on the magnitude of earthquake, and distance of a particular location from epicenter, but also on the local geology\cite{iitk4}. Hence, we considered some historical data of the reach of different earthquakes of different magnitudes. For example, the $2001$ earthquake of Bhuj with magnitude (M) $7.7$ on Richter scale affected areas up to $375$ km from its epicenter, the $1997$ earthquake of Jabalpur with M$=6.0$ had its minimum effects felt up to $275$ km\cite{niceemajor}. Hence, for simulation purpose, we consider the average maximum radial distance up to which the minimum effects of earthquakes may be realized as $100$km, $200$km, $300$km, $400$km, and $500$km, for earthquakes with $4.5<=$ M $<5.5$, $5.5<=$ M $<6.5$, $6.5<=$ M $<7.5$, $7.5<=$ M $<8.5$, and $8.5<=$ M $<9.5$, respectively. The distance up to which the earthquake may cause optical fiber link failures has been considered as half the distance of minimum realization, i.e, $50$km, $100$km, $150$km, $200$km, and $250$km, for earthquakes with $4.5<=$ M $<5.5$, $5.5<=$ M $<6.5$, $6.5<=$ M $<7.5$, $7.5<=$ M $<8.5$, and $8.5<=$ M $<9.5$, respectively, for simulation purpose.

As per NSSC, we assume a maximum searchable area for node relocation to be of $5\times5$ grids which represents an area of about $260\times260$ km. The topology obtained after node relocations using the proposed SZANR as discussed in Section III of this paper, is shown in Fig. 5. The distances between links are the minimum possible distances between each pair of nodes. However, the actual deployment of physical links may not be through the shortest distance depending on various factors such as mountains, water bodies, etc., present between the nodes.

The metrics of importance considered are earthquake survivability ratio (ESR), and bandwidth loss in THz. ESR is defined as the ratio of the number of established connections survived after earthquake and the number of connections established before earthquake. We have shown with simulations, the effects of earthquakes on the presently deployed WDM network, and the effects on future EONs if same set of earthquakes occur in future. EON technology is rapidly developing, and efforts are underway to migrate from current WDM networks to EONs in near future\cite{ruiz2014planning}. We assume that the Indian networks will be migrated to EON by year $2020$, and demonstrate the effect of earthquakes on the future EON through simulation of same set earthquakes that occurred in the past.

We perform simulation on seismic zone unaware (SZU) network, i.e., existing RailTel network topology, and SZANR, i.e., the topology obtained using the proposed node relocation scheme. Static routing and wavelength assignment (RWA)\cite{zang2000review}, and static routing and spectrum allocation (RSA)\cite{jinno2009spectrum} has been performed to establish lightpath requests in this work for WDM networks, and EON, respectively. The spectrum assignment scheme used is first-fit\cite{zang2000review}. We also compare the results by using dedicated path protection (DPP)\cite{sahas2003survivable} for all the unique lightpath requests possible in both SZU and SZANR. The frequency slot requirements of lightpath requests have been generated uniformly between $2$ and $10$ slots with guard band included.

In Fig. 7, the ESR obtained for different schemes on occurrence of each major past earthquake as given in Fig. 3, is shown. It is evident from the mean ESR graph that using SZANR, the already existing network can be made more survivable against earthquakes than using DPP in the already existing network, which requires double spectrum for protection.

\begin{figure}[!t]
\centering
\includegraphics[width=3.2in]{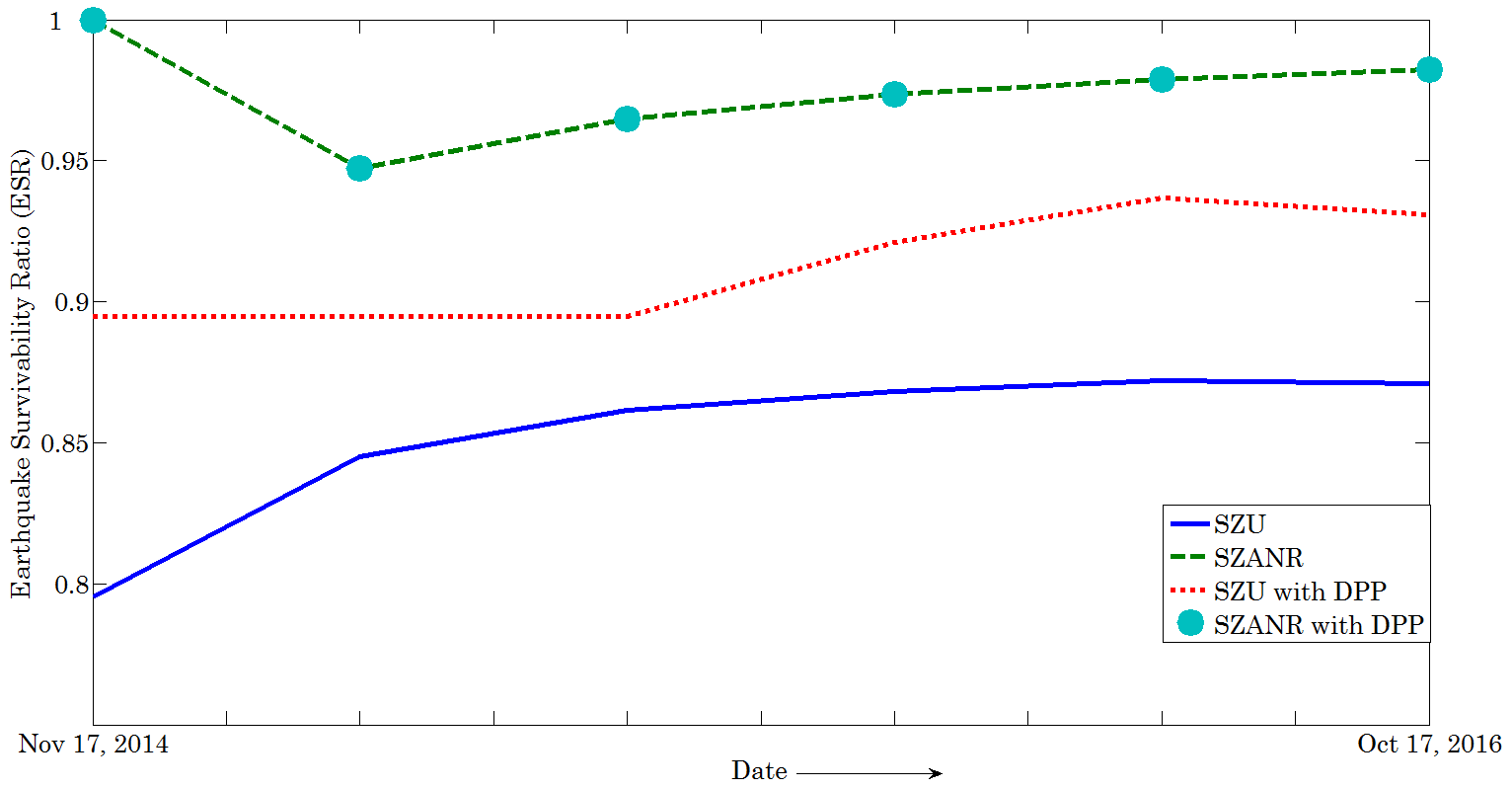}
\caption{Variation in earthquake survivability ratio (ESR) obtained in WDM optical network for SZU, SZANR, SZU with DPP, SZANR with DPP, using the information of earthquakes occurred in past two years in India.}
\label{fig_sim}
\end{figure}


\begin{figure}[!t]
\centering
\includegraphics[width=3.2in]{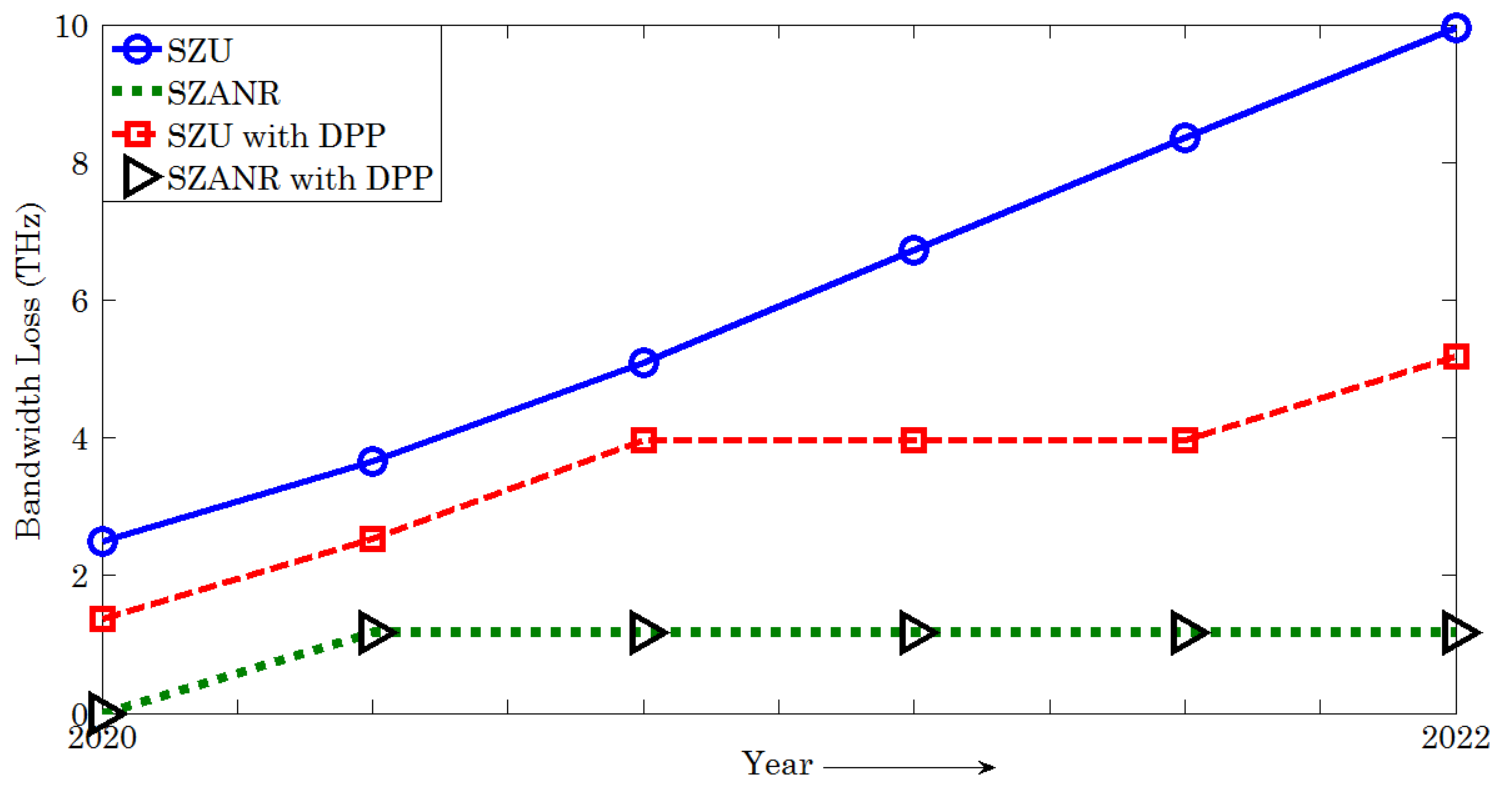}
\caption{Total bandwidth loss that may incur in future EONs if earthquakes similar to those occurred in past two years are repeated.}
\label{fig_sim}
\end{figure}

In Fig. 8, the effect of earthquakes occurred in the recent past on the SZU and SZANR networks is shown. It can be seen that the topology relocated using the proposed SZANR achieves higher ESR than both SZU alone, and SZU combined with DPP. With an envision to design the future optical networks highly robust against earthquakes, we obtain the effects of loss in bandwidth on SZU and SZANR network topologies under EON environment, if the same set of earthquakes is repeated in future. That is, if past major earthquakes are repeated in next $200$ years from $2020$ to $2220$, and past two year earthquakes are repeated in future from $2020$ to $2022$.

From Fig. 9, it is observed that if the same set of earthquakes is repeated in future, the bandwidth loss will be reduced by almost half, if existing network topology is relocated using the proposed SZANR. Fig. 10 also suggests that SZANR scheme is not limited to achieve long term benefits only, however, it can be seen that significant improvements in survivability and bandwidth loss can be achieved for a shorter duration of time too.

\section{Conclusion}

Earthquakes have resulted in severe communication network failures in the past. The frequency of natural disasters such as earthquakes is increasing with time due to global warming, tectonic changes, and other such effects. Research efforts have been made to improve the optical network survivability in events of component failures through various protection and restoration methods. With the advancements in seismological technologies, the information on disaster forecasts and vulnerable areas is becoming more accurate. Hence, along with conventional network survivability methods, the information available on seismic hazard maps of different parts of the world, can improve backbone optical network survivability to a greater extent. In this work, we proposed a scheme of node relocation (SZANR) that achieves improved survivability in events of earthquakes through slight changes in the node location of the already existing optical networks. The analysis has been made using the actual seismic hazard maps of India, earthquakes occurred in the past, and the real world Indian RailTel optical network.


\section*{Acknowledgment}

This work is supported by the Department of Electronics and Information Technology (DeitY), India.



%


%

\end{document}